\documentclass[]{aastex631}

\usepackage{comment}
\newcommand{\Gaspery}{\texttt{Gaspery}}
\newcommand{\gaspery}{\texttt{gaspery}}

\begin{document}

\title{\gaspery: Optimized Scheduling of Radial Velocity Follow-Up Observations for Active Host Stars}

\author[0000-0002-1910-5641]{Christopher Lam}
\affiliation{Center for Computational Astrophysics, Flatiron Institute, Simons Foundation, 162 Fifth Ave, New York, NY 10010, USA}
\affiliation{University of Florida Department of Astronomy, 211 Bryant Space Science Center, PO Box 112055, Gainesville, FL 32611, USA; c.lam@ufl.edu}

\author[0000-0001-9907-7742]{Megan Bedell}
\affiliation{Center for Computational Astrophysics, Flatiron Institute, Simons Foundation, 162 Fifth Ave, New York, NY 10010, USA}

\author[0000-0002-3852-3590]{Lily L. Zhao}
\affiliation{Center for Computational Astrophysics, Flatiron Institute, Simons Foundation, 162 Fifth Ave, New York, NY 10010, USA}

\author[0000-0002-5463-9980]{Arvind F.\ Gupta}
\affil{U.S. National Science Foundation National Optical-Infrared Astronomy Research Laboratory, 950 N.\ Cherry Ave., Tucson, AZ 85719, USA}

\author[0000-0002-3247-5081]{Sarah A.\ Ballard}
\affiliation{University of Florida Department of Astronomy, 211 Bryant Space Science Center, PO Box 112055, Gainesville, FL 32611, USA; c.lam@ufl.edu}

\begin{abstract}

Radial velocity (RV) follow-up is a critical complement of transiting exoplanet surveys like the {\it Transiting Exoplanet Survey Satellite} (\textit{TESS}), both for validating discoveries of exoplanets and measuring their masses. Stellar activity introduces challenges to interpreting these measurements because the noise from the host star, which is often correlated in time, can result in high RV uncertainty. A robust understanding of stellar activity and how its timescales interact with the observing cadence can optimize limited RV resources. For this reason, in the era of over-subscribed, high-precision RV measurements, folding stellar activity timescales into the scheduling of observation campaigns is ideal. We present \gaspery, an open-source code implementation to enable the optimization of RV observing strategies. \Gaspery\ employs a generalized formulation of the Fisher Information for RV time series that also incorporates information about stellar correlated noise. We show that the information contained in an observing strategy can be significantly affected by beat frequencies between the orbital period of the planet, the stellar rotation period, and the observation epochs. We investigate how the follow-up observing strategy will affect the resulting radial velocity uncertainty, as a function of stellar properties such as the spot decay timescale and rotation period. We then describe two example use cases for \gaspery: 1) calculating the minimum number of observations to reach an uncertainty tolerance in a correlated noise regime and 2) finding an optimal strategy given a fixed observing budget. Finally, we outline a prescription for selecting an observing strategy that is generalizable to different targets. 
\end{abstract}

\section{Introduction} \label{sec:intro}
Transit survey missions such as {\it Kepler} \citep{borucki_kepler_2010} and the {\it Transiting Exoplanet Survey Satellite} \citep[\textit{TESS};][]{ricker_transiting_2015,guerrero_tess_2021} have identified thousands of exoplanets \citep{nasa_exoplanet_archive_planetary_2023}. There exists an important synergy between planets identified via transit and radial velocity (RV) follow-up observations, both for authenticating transit signals and also for enabling more detailed planetary characterization.  Measurements of both the planet radius and mass enable empirical constraints on the internal structure and composition of exoplanets, allowing us to distinguish potentially terrestrial planets from icy or atmosphere-dominant worlds \citep{valencia_radius_2007, seager_mass-radius_2007, fortney_planetary_2007}.  Measuring the planet mass precisely is also key to the robust interpretation of atmospheric observations. For example, \cite{batalha_precision_2019} found that a mass precision of approximately 20\% is recommended for detailed atmospheric characterization.

Ground-based radial velocity follow-up requires significant resources, however, and given the thousands of transiting planets, they are heavily oversubscribed. Since the April 2018 launch of \textit{TESS}, which focuses on brighter, nearby stars, only 432 of the 7125 planet candidates detected by the mission have been confirmed \citep{nasa_exoplanet_archive_planetary_2023}\footnote{Accessed on 2024-03-26}. The remaining planet candidates constitute a lengthy and ever-growing backlog of targets that will benefit from the design of efficient survey schedules \citep{crass_extreme_2021}. It is therefore crucial to maximize the information yield from a limited budget of observations, which follows in the steps of informed risk assessment paradigms for other problems within exoplanet science, including the optimization of measurement uncertainties in transit observations \citep{carter_analytic_2008,price_transit_2014} and in spectrograph design \citep{figueira_radial_2016}, culminating with RV follow-up \citep{burt_simulating_2018}.

One significant challenge specific to RV follow-up is the noise resulting from stellar surface variability, which can exhibit temporal variations in ways that can mimic or confuse the periodic signal of planets \citep{aigrain_simple_2012}. This contributes to uncertainties and biases in exoplanet mass measurements. An increased volume of observations can drive down these uncertainties by providing more data to constrain the stellar activity signal, but resources for follow-up are precious, especially for targets that are key to pushing forward our understanding of exoplanets and their habitability. This provides a strong motivation for RV scheduling frameworks to incorporate stellar variability modeling \citep{cloutier_quantifying_2018}. Such a framework would be particularly relevant to the case of \textit{TESS} follow-up, where the host stars are sufficiently bright for RV observation. The orbital periods of \textit{TESS}-discovered planets are also likely to be on similar 
timescales to the stellar rotation period, particularly those in the habitable zone \citep{vanderburg_radial_2016}. For this reason, they are the most likely targets to benefit from a follow-up strategy informed by stellar activity modeling. 

One promising means of modeling stellar variability is the use of Gaussian Processes (GPs). GPs are a flexible class of probabilistic models that allow users to incorporate physically motivated domain knowledge into the modeling of the covariance between samples in some stochastic process \citep{aigrain_gaussian_2023}. \citet{dumusque_radial-velocity_2017} organized coordinated efforts to evaluate different methods for characterizing stellar noise in order to retrieve planet masses, providing a structured overview of methods for stellar noise modeling in service of uncovering the planet signal. Of these methods, GP regression has emerged as a popular tool for modeling stochastic, time-correlated stellar variability \citep{rajpaul_gaussian_2015, tran_joint_2023}. Optimal RV survey scheduling using GPs has previously been done in \citet{cloutier_quantifying_2018} and \citet{gupta_fishing_2023}. 

The ``optimal" way to conduct an observing strategy may vary, dependent upon both the specific science goal and resource constraints. In this manuscript, we define an ``optimal" strategy as resulting in the most precise measurement of the radial velocity semi-amplitude induced by the planet, $K$. In this sense, our goal is to explore the best means for minimizing $\sigma_{K}$ by survey design alone, in the presence of stellar activity and given a limited number of observations. To that end, we have developed a general optimization strategy for the scheduling of RV follow-up observations of exoplanet candidates detected by the transit technique. Our generalized approach relies upon the Fisher Information, which describes the information yield about a set of free parameters, given some noise model. Our resulting framework, called \gaspery\footnote{\url{https://github.com/exoclam/gaspery}}, is an open-source \texttt{Python} package installable from PyPI \citep{lam_gaspery_2024}. \Gaspery\ enables users to craft optimal and flexible strategies for exoplanet RV follow-up, and can be generalizable to other scheduling use cases. While we have primarily developed \gaspery\ for the case of constraining a Keplerian signal in the presence of a quasi-periodic GP noise model, the package is easily extensible to other models. 

An investigation of RV scheduling strategy (assuming the presence stellar activity) requires us to address the following questions. 

\begin{itemize}
  \item \textbf{Minimum observations:} What is the minimum number of observations required to reach a specified uncertainty tolerance on the RV semi-amplitude in the face of stellar correlated noise?
  \item \textbf{Impact of key timescales on the optimal observing strategy:} How can an RV mass measurement be affected by the time-domain parameters associated with the problem of RV follow-up observation scheduling (ie., planet orbital period, stellar rotation period, observation cadence)?
  \item \textbf{Optimal allocation of observations:} What strategy optimizes the uncertainty on the RV semi-amplitude measurement, given a fixed observing budget and target?
\end{itemize}

This manuscript is organized as follows. In Section \ref{sec:methods}, we introduce our method for evaluating observing strategies, which relies upon the Fisher Information. We also describe the application of GP kernels to model stellar variability and demonstrate the effect of correlated noise on an observer's ability to estimate an exoplanet's mass. In Section \ref{sec:analysis}, we employ Fisher Information as a means to evaluate the efficacy of different RV follow-up observing strategies for different targets, investigating each of the three questions posed above. In Section \ref{sec:comparison}, we compare our approach for estimating the uncertainty of the planet RV semi-amplitude to the uncertainty actually retrieved by fitting to synthetic data. In Section \ref{sec:discussion}, we comment on the expected performance of different classes of strategies for different targets and the limitations of \gaspery. Finally, we conclude in Section \ref{sec:conclusion}.

\section{Methods}
\label{sec:methods}
In this Section, we describe our methodology for the development of \gaspery. In Section \ref{sec:fisher-information}, we introduce the framework of the Fisher Information. Here, we apply the Fisher Information to describe the information yield of RV observing strategies. In Section \ref{sec:noise-models}, we describe our noise models. We employ Gaussian Processes as a means of mimicking the effects of stellar activity upon the RV noise, in addition to photon noise. Finally, in Section \ref{sec:gaspery}, we present the software implementation of this framework, called \gaspery.

\subsection{Fisher Information} \label{sec:fisher-information}
In order to optimize the scheduling of RV follow-up observations, we turn to the Fisher Information matrix. The Fisher Information describes the information yield about a set of free parameters, $\vec{\theta}$, contained in a set of observations, $\vec{X}$. Here, we assume that the data $\vec{X}$ are drawn from a Gaussian distribution with its mean described by a model $\vec{\mu}(\vec{\theta})$ and dispersion described by the covariance matrix $\Sigma$. $\Sigma$ may include both expected photon noise levels and correlated stellar noise. Under these assumptions, the expected Fisher information can be calculated without direct reference to the data $\vec{X}$.

In this context, we calculate the Fisher Information matrix as follows:
\begin{equation}
    I_{m,n} = {\frac{\partial \vec{\mu}}{\partial \vec{\theta_m}}}^T \Sigma^{-1} \frac{\partial \vec{\mu}}{\partial \vec{\theta_n}},
\label{eq:fisher-information}
\end{equation}
where $\vec{\mu}$ is the RV model, $\vec{\theta}$ is the vector of free parameters, and $\Sigma$ is the covariance matrix. The subscripts $m$ and $n$ denote elements in the Fisher Information matrix, which have a maximum equal to the size of $\theta$ (in other words, the number of free parameters in the mean model). 

In the case of a circular orbit for a single planet, we can use a zero-eccentricity Keplerian model that is described by the following equation: 
\begin{equation}
    \vec{\mu}(\vec{t}, K, P, T_0) = -K sin(\frac{2\pi}{P_{orb}}(\vec{t}-T_0)),
\label{eq:rv-model}
\end{equation}
where $\vec{t}$ is a vector containing the RV observation time stamps, $K$ is the RV semi-amplitude due to the planet, $P_{orb}$ is the planet's orbital period, and $T_0$ is the mid-transit time. In Equation~\ref{eq:fisher-information}, $\vec{\theta}$ is therefore a vector of just the planet's orbital parameters: $K$, $P_{orb}$, $T_0$. Accordingly, the Fisher Information matrix in this work is a 3-by-3 matrix whose diagonal elements correspond to the expected information content of $K$, $P_{orb}$, and $T_0$, respectively. Inverting the Fisher Information matrix and taking the square root of the matrix element associated with $K$ -- that is, taking the first diagonal element -- yields the expected $\sigma_K$. In this work, this is the only element of interest because $K$ is directly proportional to the planet mass. 

We note that Equation \ref{eq:fisher-information} calculates the \textit{expected} Fisher information (or $\sigma_K$), which gives the expectation value for the information content (or uncertainty) for a set of observations taken at times $\vec{t}$. After the observations have been taken, the \textit{observed} Fisher information may be calculated as the second derivative of the likelihood, as described in \citet{cloutier_quantifying_2018} and others. These two values will generally be very similar but may vary due to the random draw of the noise that is present in the specific observations in question. For the current work, we focus on the expected information as a means of quantifying how well a set of observing times will work on average for constraining the parameters of interest.

We compute the Fisher Information using a \texttt{JAX} implementation. \texttt{JAX} \citep{jax2018github} is a \texttt{Python} package that provides auto-differentiation and optimized compiling capabilities, making it useful for calculating the Jacobian in Equation \ref{eq:fisher-information}. This is especially useful in the case where the RV model is no longer analytically differentiable, such as when nonzero eccentricity is introduced (see Section \ref{sec:limitations}). 

\subsection{Noise models} \label{sec:noise-models}
Let us first consider the most basic case of a circular, Keplerian RV model with white noise only. We assume that the white noise is comprised of the photon noise and the white noise component of instrumental noise. Since with RV follow-up we are most concerned about maximizing information on $K$, the covariance matrix in the white noise case is a diagonal matrix of constant uncertainty on the measured RV signal. 

In practice, stellar variability introduces a signal that can confound or even dwarf the planet signal. 
Over the last decade, stellar variability modeling has increasingly relied on a quasi-periodic Gaussian Process (GP) kernel to describe the correlated noise from the star \citep[e.g.][and numerous references therein]{haywood_planets_2014, barragan_radial_2019, barragan_pyaneti_2022, aigrain_gaussian_2023}. This GP kernel captures two separate timescales of periodicity and takes the following form: 
\begin{equation}
    k_{t_i,t_j} = \sigma_{corr}^2 exp[-\frac{(t_i - t_j)^2}{{\tau}^2} - \frac{sin^2 \frac{(\pi(t_i - t_j))}{P_{rot}}}{{\eta}^2}],
\label{eq:correlated-noise}
\end{equation}
where $t_i$ and $t_j$ are two time stamps, $\sigma_{corr}$ is the amplitude modifier (otherwise known as the correlated noise amplitude), $\tau$ is the spot lifetime, $P_{rot}$ is the rotation period of the star, and $\eta$ is the harmonic complexity (also known as the Gaussian Process smoothing parameter). We use this formulation to populate the covariance matrix, $\Sigma$ (with dimensions equal to the length of the observing strategy time series), in order to calculate the Fisher Information (see Equation \ref{eq:fisher-information}) for the correlated noise case.

\subsection{\Gaspery}
\label{sec:gaspery}
To facilitate the use of Fisher Information to evaluate and discern between RV follow-up strategies, we developed a \texttt{Python} package, called \gaspery. From an input of the planet and stellar parameters, as well as a list of time stamps constituting the observing strategy, \gaspery\ calculates an expected Fisher Information and corresponding $\sigma_K$. \Gaspery\ relies upon \texttt{JAX} \citep{jax2018github} to calculate the Fisher Information and to compute the Jacobian of the mean model. It employs \texttt{tinygp} \citep{foreman-mackey_dfmtinygp_2024} to construct a Gaussian Process kernel representing the noise model. \texttt{JAX}'s auto-differentiation capabilities allow us to quickly take the partial derivatives that comprise the Jacobian in Equation \ref{eq:fisher-information}, while \texttt{tinygp} allows us to populate the covariance matrix in Equation \ref{eq:fisher-information}. \Gaspery\ is downloadable from PyPI, and installation instructions and the code base can be found at \url{https://github.com/exoclam/gaspery}.

\section{Quantifying Performance With Fisher Information}
\label{sec:analysis}

We apply \gaspery\ as a tool in this paper to answer the three questions outlined in Section \ref{sec:intro}. First, what is the minimum number of observations required to achieve an uncertainty tolerance for $K$ (Section \ref{sec:au-mic})? Second, how do the planet period and stellar variability timescales affect the choice of optimal observing strategy (Sections \ref{sec:time_sampling}, \ref{sec:beat-frequencies}, and \ref{sec:stellar-parameters})? Third, what is the optimal allocation of observations over the course of an observing window, given a fixed budget of observations (Section \ref{sec:fixed-budget})?

\subsection{Minimum observations}
\label{sec:au-mic}
AU Mic is a young (22 Myr old), active early M-dwarf that is host to at least two planets \citep{wittrock_validating_2023} and has been the target of extensive RV campaigns from many different instruments \citep{plavchan_planet_2020, klein_investigating_2021, cale_diving_2021}. Its young age and high level of stellar correlated noise make it an archetypal example of a planetary system that requires both a large quantity of observations and high quality stellar activity modeling in order to meaningfully characterize the masses of its planets. We make use of a slightly modified version of the AU Mic planetary system in order to illustrate how a principled approach based on the Fisher Information can enable the design of a realistically optimal observing strategy. In this simplified case study, the host star has only one planet on a circular orbit. The stellar noise properties and the planetary orbital period are set by the \citet{klein_investigating_2021} study of AU Mic b. The
se properties are listed in Table \ref{tab1}. Throughout the text that follows, we refer to an ``AU Mic-like star" and an ``AU Mic b-like planet" as a star and planet that have these parameters from \citet{klein_investigating_2021}.

\begin{figure}
\includegraphics[width=.9\textwidth]{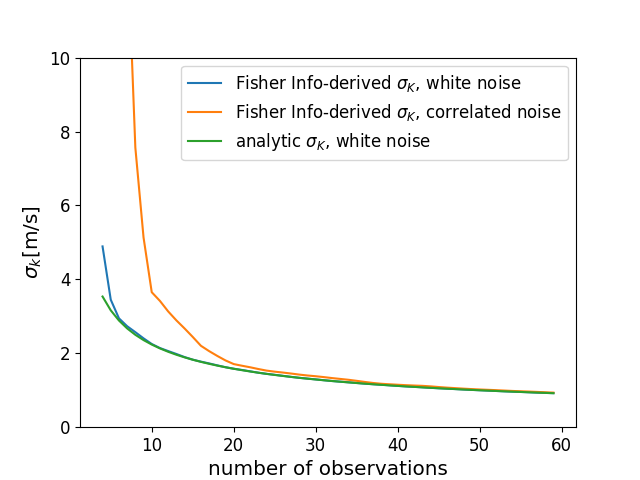}
\caption{The uncertainty on the planet RV semi-amplitude, $\sigma_K$, as a function of the number of observations of a single AU Mic b-like planet transiting an AU Mic-like star, under a white noise regime (blue) and correlated noise regime (orange). We fix the observing cadence at one observation per night, every night. For reference, we overlay the Fisher Information-derived $\sigma_K$ curves against the analytic $\sqrt{2/\textrm{N}}$ relation between number of observations and $\sigma_K$ (green).}
\label{fig:n_obs}
\end{figure}

In Figure \ref{fig:n_obs}, we show that $\sigma_K$ as calculated using our Fisher Information formulation in a regime without correlated noise (blue) agrees with the analytic $\sqrt{2/N}$ relation given by \citet{cloutier_quantifying_2018} (green), where $N$ is the number of observations. In order to marginalize over phase dependence, we randomly sample the observing start time up to one planet orbital period from a fiducial start time and average the curves over 100 draws. For the Fisher Information-derived curves, we use a white noise amplitude of 5 m/s. We show that in the white noise-only case, as the Fisher Information for $K$ grows with more measurements, the uncertainty on $K$ decreases as $\sqrt{2/N}$. 

In the case of correlated noise, the amount of excess noise introduced depends on the observing cadence used. This effect is largest at small $N$, and we will show in Sections \ref{sec:time_sampling} through \ref{sec:fixed-budget} that in this regime, efficient follow-up depends on minimizing the influence of correlated noise. As we also show in Fig \ref{fig:n_obs}, injecting correlated noise (orange) significantly increases the uncertainty at lower numbers of observation but converges with the white noise case after about 30 observations. For our correlated stellar noise model, we use the quasi-periodic GP kernel parameters from \citet{klein_investigating_2021}, which are \{$P_{rot}$ = 4.84 days, $\tau$ = 100 days, $\sigma_{corr}$ = 43 m/s, $\eta$ = 0.4\} (Table \ref{tab1}).

\begin{figure}
\includegraphics[width=.5\textwidth]{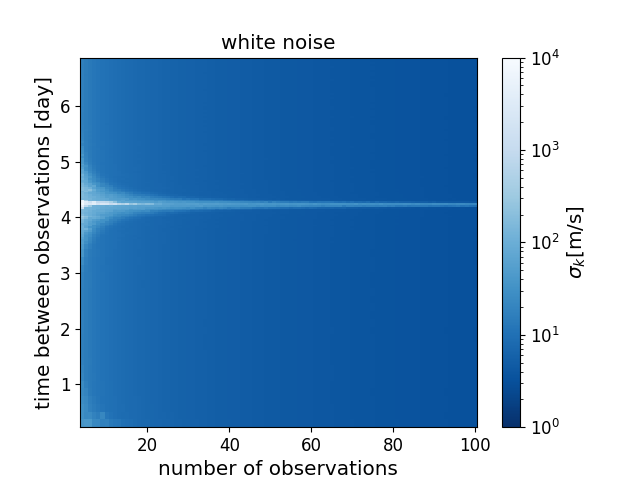} 
\includegraphics[width=.5\textwidth]{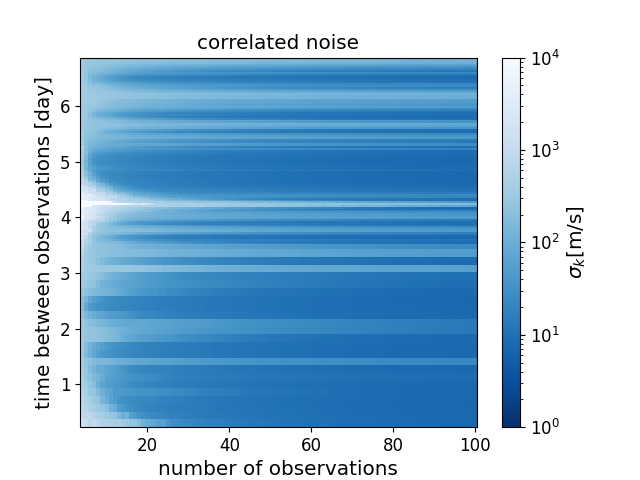}
\caption{Sensitivity maps showing how the uncertainty on the planet RV semi-amplitude, $\sigma_K$, changes for different combinations of number of observations (x-axis) and observing cadence (y-axis). The color bar is log-scaled in order to highlight the structure in these plots. Left: In the white noise case, the uncertainty on K decreases gradually as the number of observations increases. Additionally, there is a band of high uncertainty for an observation cadence equal to half of the planet's orbital period. Right: In the correlated noise regime, other observing cadences also give rise to high uncertainties at beat frequencies between the planet orbital period and modes of the stellar rotation period.}
\label{fig:n_obs_vs_cadence}
\end{figure}

In Figure \ref{fig:n_obs_vs_cadence}, we illustrate how an active star can impinge on the expected $\sigma_K$ of a given observing strategy, using a side-by-side comparison of two sensitivity maps showing the expected $\sigma_K$ for different combinations of observing cadence and number of observations of AU Mic b. The left panel corresponds to an idealized observing scenario of AU Mic b in a white noise regime, while the right panel corresponds to a correlated noise regime. From Figure \ref{fig:n_obs}, we would expect $\sigma_K$ to decrease with more observations. We see this reflected in the background color gradient, which changes from higher uncertainty on K (lighter blue) to lower uncertainty on K (darker blue) along the x-axis. The more obvious feature in the white noise case, however, is the band of high $\sigma_K$ at an observing cadence equal to half of the orbital period of the planet, which suggests that physical properties of the target system -- in this case, $P_
{orb}$ -- can inform what observation strategies should be avoided. 

The right-hand plot in Figure \ref{fig:n_obs_vs_cadence} shows that introducing a realistic correlated noise model produces additional lines of higher uncertainty at beat frequencies between the planet orbital period and modes of the stellar rotation period. In general, the right panel of Figure \ref{fig:n_obs_vs_cadence} indicates that adding stellar correlated noise to a set of RV observations penalizes some observing strategies more than others in a way that depends on both the system's planet and stellar properties. We investigate how these properties lead to poor observing strategies in the following three subsections.

\begin{table}
  \centering
  \caption{Stellar Parameters of AU Mic and Planet Parameters of AU Mic b from \citet{klein_investigating_2021}}
    \begin{tabular}{ c|c }
    GP parameter & GP prior \\ 
     \hline
     $\theta_{corr}$ & $43^{+11}_{-8}$ m/s \\  
     $\tau$ & 100 days \\  
     $P_{rot}$ & 4.84 $\pm{0.01}$ days \\  
     $\eta$ & 0.4 \\  
     \hline
     K & $8.5^{+2.3}_{-2.2}$ m/s \\
     $P_{orb}$ & 8.46321 $\pm{0.00004}$ days \\
     $T_0$ & 2458651.993 $\pm{1}$ [BJD] \\
     \hline
     $\theta_{WN}$ & 5 m/s \\  
    \end{tabular}  
  \label{tab1}
\end{table}

\subsection{Optimizing time sampling of observations} \label{sec:time_sampling}

While the parameter space of RV follow-up observation strategies is broad and complex, it is simplest and most practical to start with the basic case of evenly spaced observations. Within this constraint, we can explore how different observing cadences perform as a function of the star and planet properties. By plotting the expectation value of $\sigma_K$ for different combinations of planet orbital period and observing cadence, we can study whether and how different time-based variables can interact to potentially reduce the information content of a set of observations of a system. Figure \ref{fig:cadence_vs_period} shows a sensitivity map similar to Figure \ref{fig:n_obs_vs_cadence}, except here we fix the number of observations at 30, vary the observing cadence along the x-axis, and vary the orbital period of the planet along the y-axis on a log scale from 1 to 100 days. We continue to use a correlated noise model with AU Mic-like stellar parameters. We trace out the compl
ex features evident on the map by overlaying analytic curves and discuss them in detail in Section \ref{sec:beat-frequencies}. We then explore how this structure changes with different stellar parameters in Section \ref{sec:stellar-parameters}.

\subsection{Key frequencies} \label{sec:beat-frequencies}

\begin{figure}
\includegraphics[width=1.\textwidth]{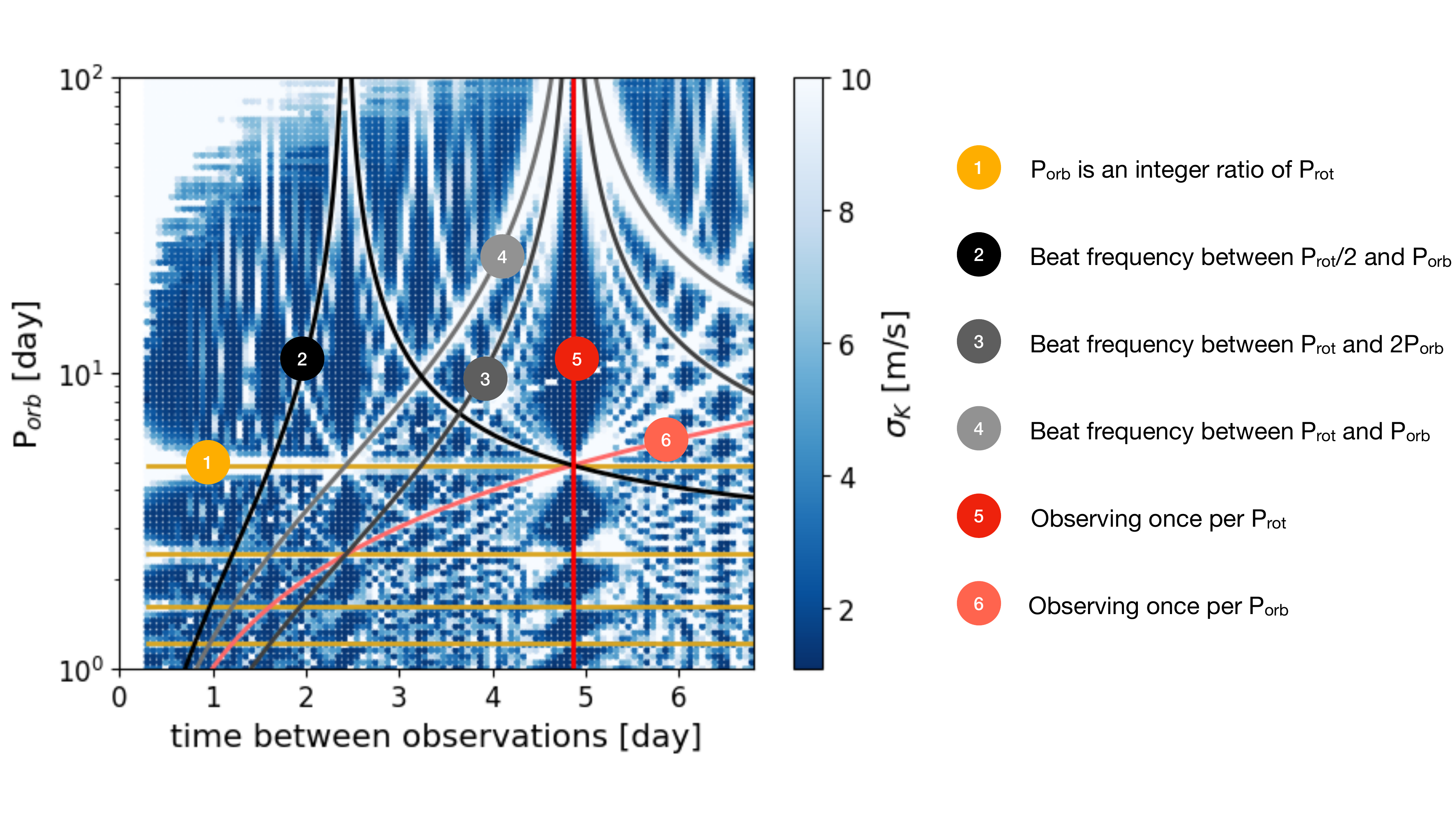}\caption{Sensitivity map showing the uncertainty on the RV semi-amplitude, $\sigma_K$, for different combinations of orbital period and observing cadence. We overplot analytic curves and labels corresponding to key frequencies that hinder our ability to measure $K$, which are described in Section \ref{sec:beat-frequencies}.}
\label{fig:cadence_vs_period}
\end{figure}

Several key frequencies shape the major features that we see in Figure \ref{fig:cadence_vs_period}. One of these is the under-sampling of the planet signal resulting from observing at a cadence equal to either $P_{orb}$ (the labeled horizontal yellow line, denoted by (1)) or a small integer ratio of $P_{orb}$ (the other horizontal yellow lines). These cases repeatedly sample the same part of the phase curve, resulting in poor phase coverage and a large $\sigma_K$. 

Another key feature in Figure \ref{fig:cadence_vs_period} results from periodic starspots, which can give rise to beat frequencies that mimic the planet signal, hindering our ability to measure the planet mass. A beat frequency occurs when two or more periodicities interact to form an interference pattern. In the case of RV follow-up, beat frequencies can be produced in a number of different ways, resulting from interactions between: 
\begin{itemize}
    \item the observing cadence and the orbital period,
    \item the orbital period and the rotation period,
    \item the observing cadence and the rotation period,
    \item and all three variables.
\end{itemize}. A beat frequency between the orbital period ($P_{orb}$) and the rotation period ($P_{rot}$), for example, is calculated as:
\begin{equation}
   f = \frac{1}{\frac{1}{P_{orb}} \pm \frac{1}{P_{rot}}}.
\label{eq:beat-frequency}
\end{equation}
A beat frequency between the observing cadence ($P_{obs}$), orbital period, and stellar period would look like: 
\begin{equation}
   f = \frac{1}{\frac{1}{P_{orb}} \pm \frac{1}{P_{rot}} \pm \frac{1}{P_{obs}}}.
\label{eq:beat-frequency-all}
\end{equation}
The spot lifetime is another (quasi)-periodic variable, but since we have set $\tau$ for our fiducial system to be much longer than the other periodic variables considered in this exercise (and most relevantly, $P_{rot}$), it is less significant of a consideration other than as a modulator for the shape of the contours in Figure \ref{fig:cadence_vs_period}. We discuss $\tau$ in further detail in Section \ref{sec:stellar-parameters}.

Since Equations \ref{eq:beat-frequency} and \ref{eq:beat-frequency-all} each have a plus and minus sign, they must generally trace \textit{two} distinct curves. In observing cadence-orbital period space, these two curves produce a ``circus tent" shape that asymptotically meets at the relevant integer ratio of the stellar rotation period. For example, in Figure \ref{fig:cadence_vs_period}, the beat frequency between $P_{rot}$/2 and $P_{orb}$ (marked by the black (2) circle) has a left component characterized by the additive version of Equation \ref{eq:beat-frequency} and a right component characterized by the subtractive version of Equation \ref{eq:beat-frequency}. In Figure \ref{fig:cadence_vs_period}, we plot only positive values from the subtractive versions of our beat frequency equations. The three circus-tent shaped contours denoted by black and gray lines in Figure \ref{fig:cadence_vs_period} correspond to the following beat frequencies and show that there are many ways
 the stellar rotation and planet orbital periods can interact to produce unfavorable observing strategies: (2) corresponds to the beat frequency between half of $P_{rot}$ and $P_{orb}$; (3) corresponds to the beat frequency between $P_{rot}$ and two times $P_{orb}$; and (4) corresponds to the 1:1 beat frequency between $P_{rot}$ and $P_{orb}$. 

\begin{figure}
\includegraphics[width=.75\textwidth, height=6.5cm]{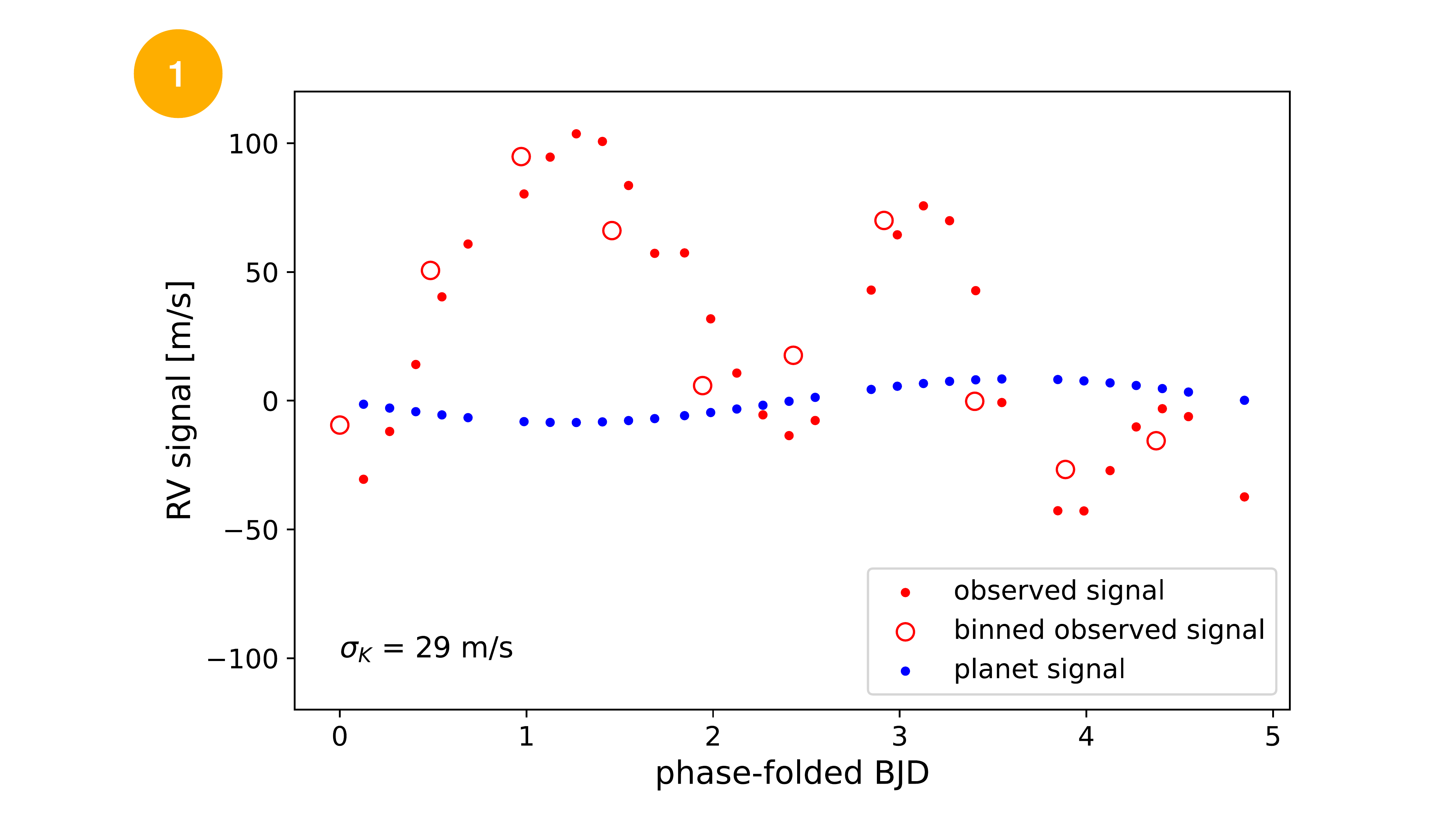}
\includegraphics[width=.75\textwidth, height=6.5cm]{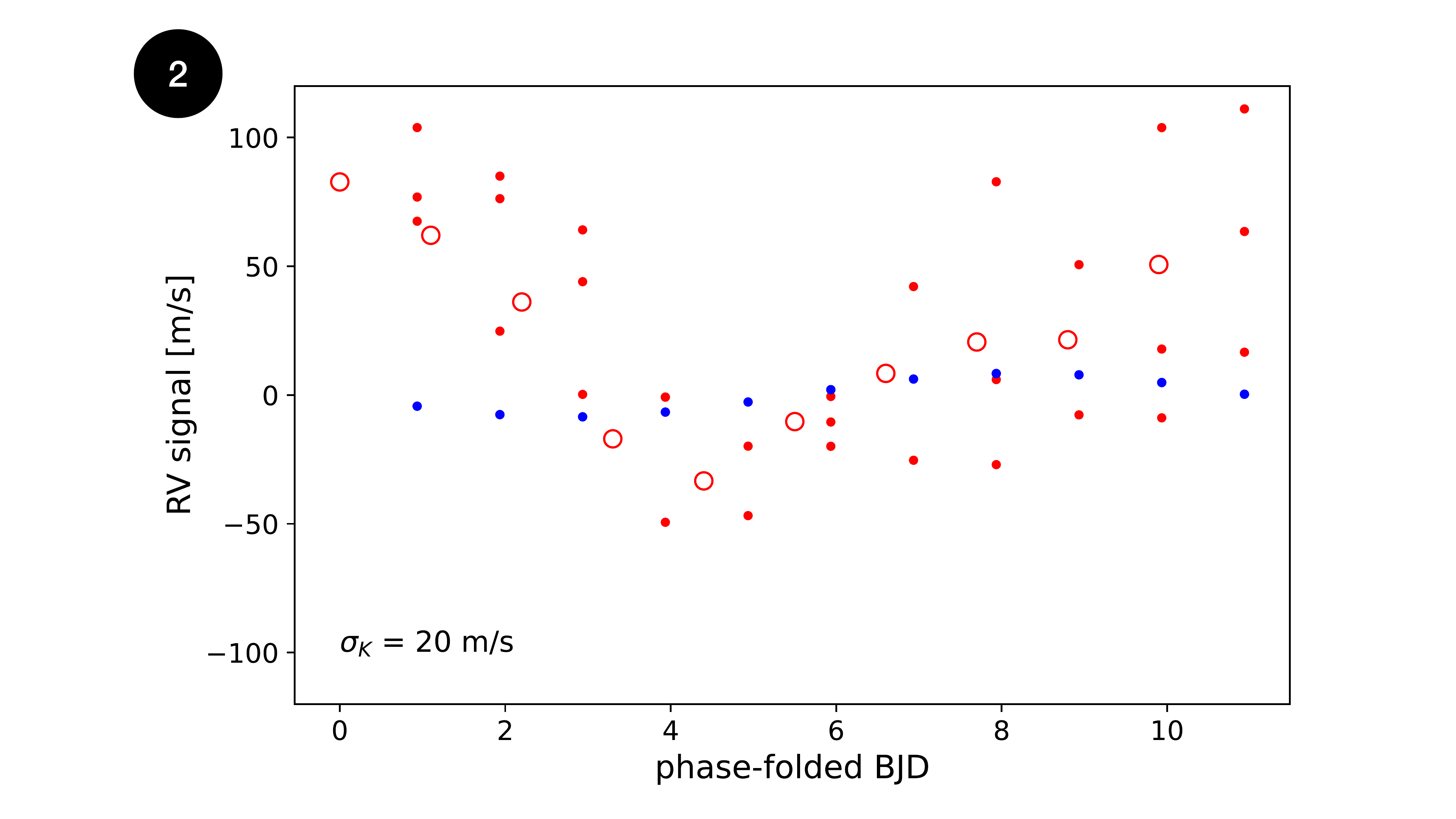}
\includegraphics[width=.75\textwidth, height=6.5cm]{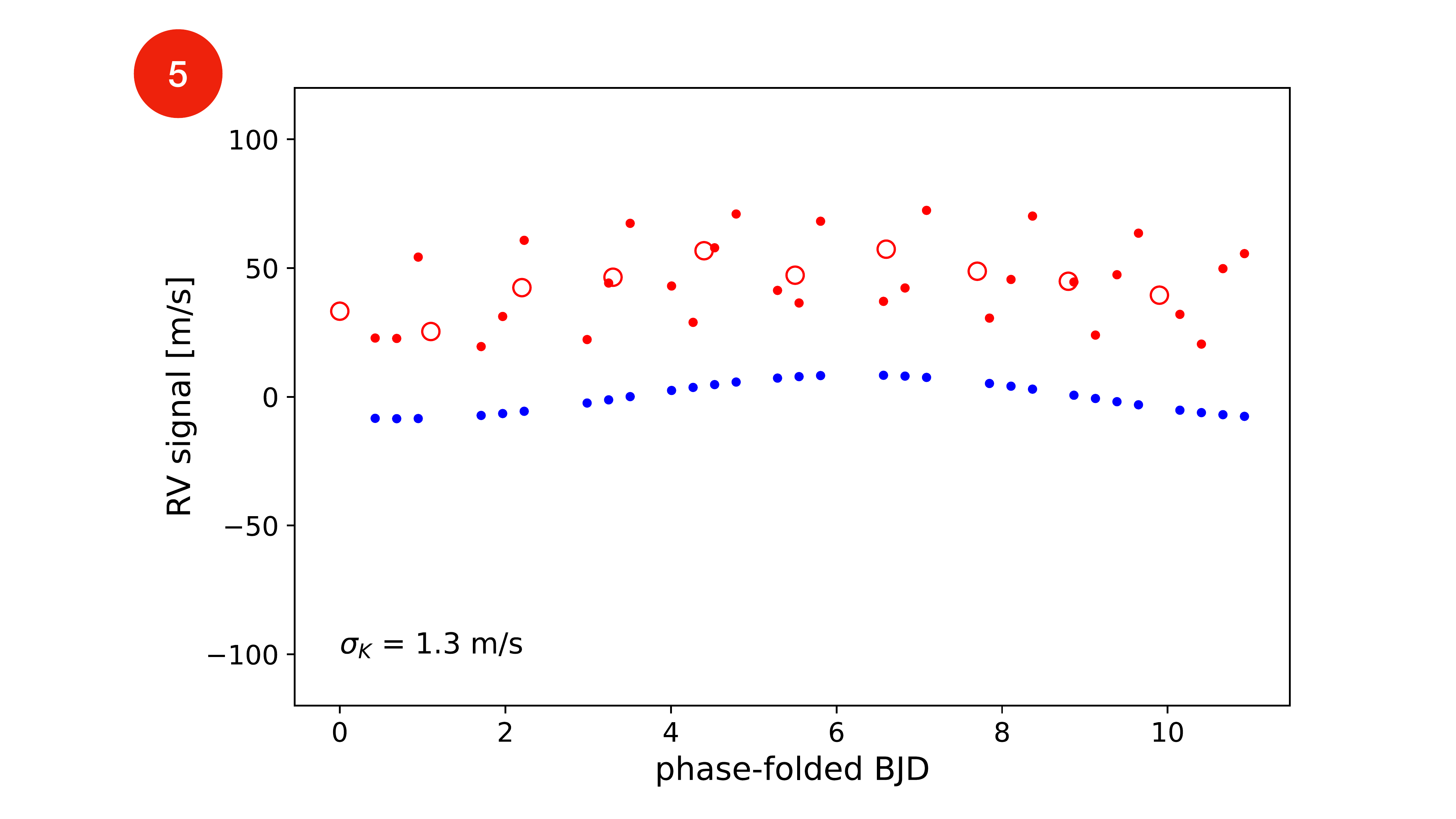}
\caption{Phase-folded RV curves showing three different scenarios of observation cadence and planet orbital period combinations. Blue points indicate the sampled injected planet signal. Red filled-in points indicate the sum of the sampled stellar signal and the planet signal. Red open circles indicate the binned stellar plus planet signal, where bin width is one tenth of the $P_{orb}$, such that there are ten bins. Each panel is associated with a different location on Figure \ref{fig:cadence_vs_period}. \textbf{Top:} Suboptimal case with $P_{orb}$ = $P_{rot}$ ((1) in Figure \ref{fig:cadence_vs_period}). \textbf{Middle:} An observation strategy on a beat frequency curve ((2) in Figure \ref{fig:cadence_vs_period}). \textbf{Bottom:} When the observing cadence is equal to $P_{rot}$ and $P_{orb}$ is not equal to $P_{rot}$ ((5) in Figure \ref{fig:cadence_vs_period}), making the sampled stellar noise effectively offset from the planet signal.}
\label{fig:beat-frequency-scenarios}
\end{figure}

To demonstrate how poor information content from an observing strategy manifests in simulated RV observations, we choose three of the scenarios highlighted in Figure \ref{fig:cadence_vs_period} and plot their phase-folded simulated data in Figure \ref{fig:beat-frequency-scenarios}. The stellar data is simulated using a quasi-periodic Gaussian Process kernel evaluated at each time stamp, while the planet data is simulated using the Keplerian mean model evaluated at each time stamp. The observed signal is the sum of these two time series, perturbed by a white noise term of amplitude 5 m/s. Observing a planet with $P_{orb}$ equal to $P_{rot}$ ((1) in Figure \ref{fig:cadence_vs_period}) is illustrated in the top phase plot in Figure \ref{fig:beat-frequency-scenarios} and leads to a poor $\sigma_K$ of 29 m/s. Sampling along the beat frequency between $P_{rot}$ and $P_{orb}$ ((2) in Figure \ref{fig:cadence_vs_period}) is shown in the middle phase plot in Figure \ref{fig:beat-frequency-scenarios} and also leads to a poor $\sigma_K$ of 20 m/s. For comparison, we also present in the bottom panel of Figure \ref{fig:beat-frequency-scenarios} a scenario in which a target is observed at a cadence equal to the stellar rotation period ((5) in Figure \ref{fig:cadence_vs_period}), while $P_{orb}$ is not one of the failure modes delineated in yellow in Figure \ref{fig:cadence_vs_period}. This specific strategy samples the same face of the star each time, representing a special case in which the information about the stellar signal is limited, but the absence of a beat frequency between $P_{orb}$ and $P_{rot}$ allows the stellar variability to be treated as an offset from the planet RV signal. This allows the expected value of $\sigma_K$ for this latter scenario to be much smaller than the previous two scenarios, with $\sigma_K$ = 1.3 m/s (we further discuss this strategy archetype in Section \ref{sec:comparison}). While this is one way to keep $\sigma_K$ low, the more gen
 eral principle demonstrated by Figure \ref{fig:cadence_vs_period} is that the best observing strategies avoid beat frequencies from stellar variability while also maximizing phase coverage of the planet's RV signal. 

Figure \ref{fig:cadence_vs_period} assumes that observations can be taken at any time of day, allowing the observing cadence to vary continuously. We show the results with continuous spacing between observations rather than discrete space in order to more clearly illustrate the contours of the beat frequency features. For realistic applications, in which observations have to be taken at night and are therefore separated by intervals that are closer to integer days, a discrete picture is presented in Section \ref{sec:fixed-budget}.

\subsection{Stellar parameters} \label{sec:stellar-parameters}
Beyond the fiducial target examined thus far, we also analyzed how changing the parameters of the correlated stellar signal affects the optimal observing strategy. In general, turning any knob in one direction either changes the stellar noise coherence, amplitude, or timescale -- any combination of which can impact the appropriate observing strategy for maximizing information content. In some cases, the GP hyperparameter values result in stellar noise that is so incoherent, low, or slowly-evolving as to smear out the regions of good and bad observing strategies. 

\begin{figure}
\includegraphics[width=.5\textwidth]{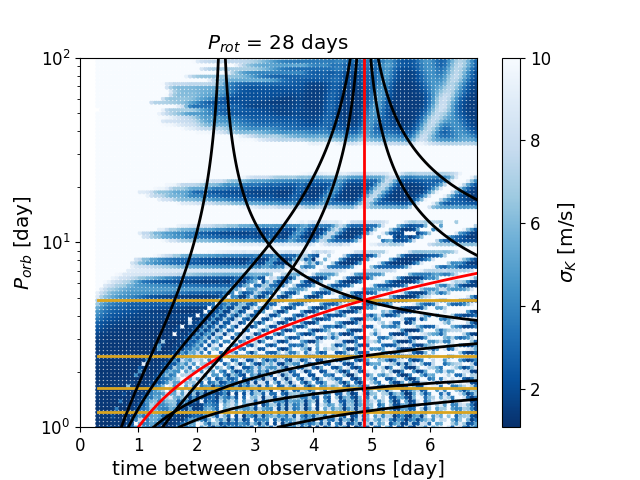}
\includegraphics[width=.5\textwidth]{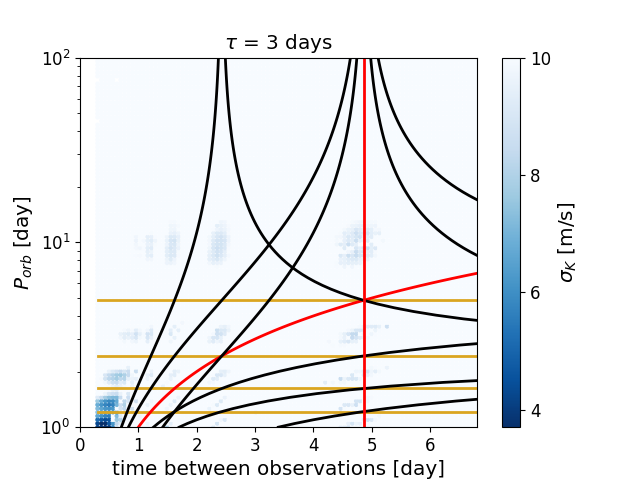}
\includegraphics[width=.5\textwidth]{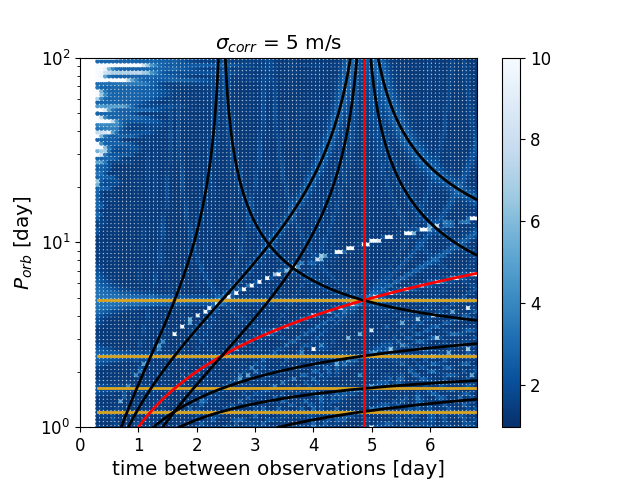}
\includegraphics[width=.5\textwidth]{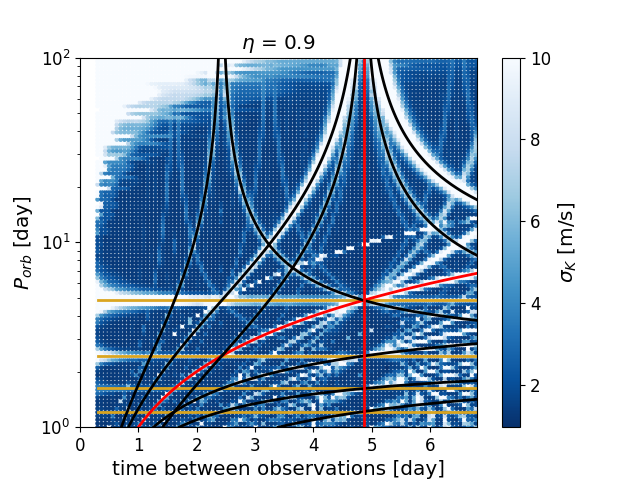}
\caption{Sensitivity maps, identical to Figure \ref{fig:cadence_vs_period}, but here we vary each of the quasi-periodic GP kernel hyperparameters, which produces changes in key features from Figure \ref{fig:cadence_vs_period}. For each plot, only one free parameter is changed from our fiducial system's values: \{$P_{rot}$ = 4.84 days, $\tau$ = 100 days, $\sigma_{corr}$ = 43 m/s, $\eta$ = 0.4\} (see Table \ref{tab1}). Drawn contours are left from Figure \ref{fig:cadence_vs_period} for ease of comparison. The \textbf{top left} panel shows how the sensitivity map changes when $P_{rot}$ is increased from 4.84 days to 28 days. The \textbf{top right} panel shows how this picture changes for a much shorter spot lifetime, when $\tau$ is decreased from 100 to 3 days (specifically, shorter than $P_{rot}$). The \textbf{bottom left} panel shows what happens when $\sigma_{corr}$ is decreased from 47 m/s to 5 m/s, which is more commensurate with the planet RV semi-amplitude. Finally, the \
textbf{bottom right} panel shows how increasing $\eta$ from 0.4 to 0.9 lowers the complexity of the correlated noise from Figure \ref{fig:cadence_vs_period}.
}
\label{fig:stellar-parameters-4}
\end{figure}

In Figure \ref{fig:stellar-parameters-4}, we illustrate how changing each of the four parameters governing the stellar activity model -- the stellar rotation period ($P_{rot}$), the spot lifetime ($\tau$), the correlated noise amplitude ($\sigma_{corr}$), and the smoothing parameter ($\eta$) -- affects the sensitivity map for different combinations of observing cadence and planet orbital period. The top left panel in Figure  \ref{fig:stellar-parameters-4} corresponds to a change in $P_{rot}$ from 4.84 days to 28 days, while holding constant all other hyperparameters from Table \ref{tab1}. We find that this change pushes the circus-tent-shaped beat frequency curves to longer cadences, from small integer ratios of 4.84 days to those of 28 days. Similarly, the horizontal lines denoting poor phase coverage move up from integer ratios of 4.84 days to integer ratios of 28 days. These low information content contours of under-sampled planet signal are also broader at this higher ste
llar rotation period. Turning the $P_{rot}$ knob therefore appears to tune the timescale (and, to some extent, the coherence) of the stellar noise. Finally, increasing the stellar rotation period pushes the region of low information content in the upper left corner of each of the plots in Figure \ref{fig:stellar-parameters-4} outward to longer cadences.

Decreasing the spot lifetime, $\tau$, from 100 days to 3 days, meanwhile, appears to tune the stellar noise coherence exclusively (see top right panel of Figure~\ref{fig:stellar-parameters-4}). We find that at lower $\tau$, the beat frequency and aliasing lines are smeared out. Specifically, when these timescales are lower than the rotation period of the star (that is, when the spot evolution occurs within the time it takes for the star to rotate), our overall ability to capture a stable picture of the star's spot activity degrades. Turning down the coherence timescale makes the spot model less stable, making it more difficult to resolve and model aliasing from beat frequencies. In the case of $\tau < P_{rot}$, every rotation period effectively bears an entirely new picture of the star because the spots keep changing so quickly. Therefore, aliases do not stand out as much because the information content is consistently poor everywhere when spots are not coherent. On the other
 hand, $\tau > P_{rot}$ bears more chances to develop a stable picture of the star. In the regime of longer-lived spots, longer-term signal coherence means that there is a significant distinction between an optimal and sub-optimal observing cadence. 

In the lower left panel of Figure \ref{fig:stellar-parameters-4}, we illustrate the effects of tuning the stellar correlated noise amplitude, $\sigma_{corr}$. Turning $\sigma_{corr}$ down from 47 m/s to 5 m/s crosses the threshold from a regime in which the stellar noise is large enough to produce distinctive features in the sensitivity map, to one in which the aliasing and beat frequency features are not strong enough to meaningfully affect the choice of observing cadence. For reference, the RV semi-amplitude, $K$, of AU Mic b is reported by \citet{klein_investigating_2021} as $8.5^{+2.3}_{-2.2}$ m/s. This suggests that the correlated noise amplitude only begins to negatively affect $\sigma_K$ when it is much larger than $K$. 

In the lower right panel of Figure \ref{fig:stellar-parameters-4}, we show that increasing the smoothing parameter smooths over all but the strongest beat frequency curves, while also weakening or completely smoothing over the harmonics of $P_{orb}$ = $P_{rot}$. Therefore, a stellar variability model with high $\eta$ (which can also be described as the inverse of the harmonic complexity) will have fewer key frequencies to consider. We caution that the smoothing parameter, $\eta$, is the least physically motivated GP hyperparameter \citep{nicholson_quasi-periodic_2022}.

There are two features that persist even at zero $\sigma_{corr}$. The region of low information content in the upper left corner of each of these plots can be explained by longer periods receiving insufficient phase coverage over 30 observations at such low cadences. The contours of low information content sweeping across the bottom half of each of these plots represent the case in which the observing cadence is equal to the orbital period of the planet (or an integer ratio or multiple of $P_{orb}$), which would be a worst case strategy independent of the star (salmon-colored contour in Figure \ref{fig:cadence_vs_period}, denoted by (6)). This also explains why toggling $\tau$, $\eta$, and $P_{rot}$ fails to change the location or morphology of these particular contours.

\subsection{Optimal allocation of observations}
\label{sec:fixed-budget}
Thus far, we have assumed that observing strategies are necessarily evenly spaced in order to explore and illustrate how stellar variability can interfere with detecting the planet signal. In this section, we introduce as part of \gaspery\ a flexible framework that allows for the construction of more general observing strategies. Besides user-inputted custom time stamps, we also enable:
 \begin{itemize}
  \item Variable number of consecutive nights of observation
  \item Variable baseline of observation (ie. the time between the last and first nights of observation)
  \item Random dropout of nights to simulate bad weather and other unplanned causes of cancelled observations
  \item User-defined nights with no observations
  \item Multiple observations per night.
\end{itemize}

\begin{figure}
\includegraphics[width=.9\textwidth]{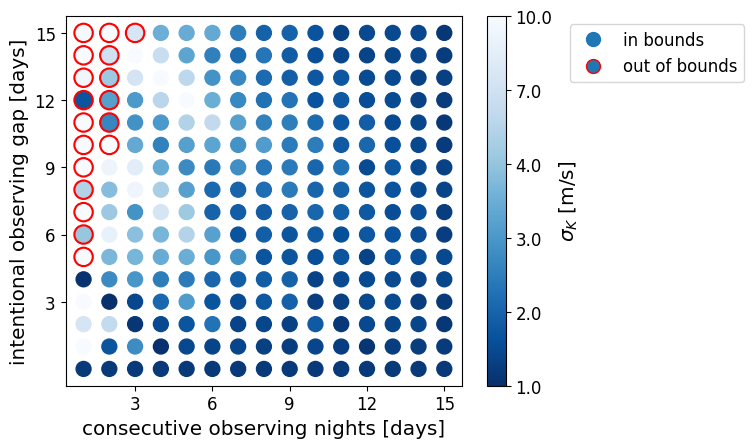}
\caption{Sensitivity map for a given fixed budget of 30 observations of an AU Mic b-like planet around an AU Mic-like star. The x-axis denotes the number of consecutive days of observation, and the y-axis denotes the number of consecutive days not observing. Strategies are outlined in red if they finish outside the 2023A NEID observing season. The colorbar is normalized to a logscale in order to better illustrate differences among strategies.}
\label{fig:fixed-budget}
\end{figure}

Figure \ref{fig:fixed-budget} illustrates an example use of \gaspery\ for designing an observing strategy for a single AU Mic b-like planet in a circular orbit around an AU Mic-like star (see Table~\ref{tab1} for exact parameters), given 30 observations during an observing semester. With this fixed observing budget, we can build many different strategies by combining two variables: the number of consecutive nights with observations and the number of nights between bursts of observations. We evaluate many possible observing strategies in Figure \ref{fig:fixed-budget}, where the x-axis denotes the number of consecutive nights observed, and the y-axis denotes the number of consecutive nights in an observing gap, during which no observations are taken. Together, they prescribe an alternating pattern of ``on" and then ``off" nights, in order to extend a finite number of observations to span a longer baseline. For example, the strategy located at plot coordinates \{3, 2\} prescribe
s observing for three consecutive days, taking a break for two days, observing another three consecutive days, taking another two-day break, and so on, until 30 observations are made, for a baseline of 48 days. For each strategy, \gaspery\ calculates the Fisher Information, and we color the corresponding point in Figure~\ref{fig:fixed-budget} by the expected $\sigma_K$.

In our example case of an AU Mic b-like planet, the simplest strategy of observing once per day every day yields the lowest expected value for $\sigma_K$. In reality, one would have to work around additional limitations on when observing can happen. As a representative case, we looked at the NEID 2023A observing season, which spanned 172 nights. In Figure~\ref{fig:fixed-budget}, we outline points in red for observing strategies that are \textit{not} executable within the bounds of the NEID 2023A observing season. Any strategy that spans more than 172 days in total baseline will be disallowed for going beyond the length of the observing semester, which generally blocks off the upper left quadrant of the figure. Additionally, specific dates within the semester that are blocked for proprietary or engineering time will force a strategy that runs into them to skip over the night and proceed longer in order to reach the 30 observation total. This leads to some strategies (such as {
8 nights on, 7 nights off}) going out of bounds, while strategies with longer baselines with no skipped dates (such as {8 nights on, 9 days off}) remain within bounds of the observing semester. This is reflected in the irregularly jagged shape of the boundary between outlined and non-outlined strategies in Figure \ref{fig:fixed-budget}. The bottom row of Figure \ref{fig:fixed-budget}, with no observing gap between consecutive nights, is redundant in that every strategy is simply observing once per night for 30 nights.\footnote{The tutorial for this case study can be found at \url{https://github.com/exoclam/gaspery/blob/main/tutorials/fixed_budget.ipynb}.}

\section{Validating gaspery's predictions with an injection-recovery study}
\label{sec:comparison}

In this Section, we validate \gaspery's predictions for observing strategy performance with a suite of injection-and-recovery tests. Ideally, our predictions for planet RV semi-amplitude uncertainty, given an observing prescription from \gaspery, ought to correlate with the measured uncertainty from a synthetic data set. We aim for these validation experiments, in addition to providing proof-of-concept, to concretely illustrate the impact of observing strategy on the final results.

We first simulate synthetic ground truth data for an AU Mic b-like planet using the orbital parameters from Table \ref{tab1}; as in the previous sections, we assume no additional planets and a circular orbit. We likewise simulate a ground truth stellar signal using \texttt{tinygp} to express a quasi-periodic GP correlated noise kernel, populated by the stellar hyperparameters from Table \ref{tab1}. The ground truth stellar and planetary signals are the same across all simulated observing strategies.

We investigate different observing strategies, where we define ``observing strategy" as the sampling of the model at prescribed time intervals. In this sense, each ``observing strategy" results in a different set of synthetic observations, though the underlying model is the same. We construct the injection-and-recovery experiment under the assumption that the stellar hyperparameters are known, and fit only the planet parameters using \texttt{numpyro} \citep{bingham_pyro_2018, phan_composable_2019}. Generating the injected planet model requires assigning to a zero-eccentricity Keplerian a set of $\{K, P, T_{0}\}$, which we draw in the following way. We sample \textit{K} from a Gaussian distribution truncated at 0 m/s and centered at 10 m/s with a spread of 20 m/s, \textit{P} from a Gaussian distribution with mean and standard deviation equal to the values listed in Table \ref{tab1}, and \textit{$T_0$} from a Gaussian distribution with mean equal to \textit{$T_0$} from Table \r
ef{tab1} and a spread of 1 hour (assuming that transit data has already tightly constrained \textit{$T_0$}). We assume white noise of 5 m/s, consistent with the rest of this work. After generating the planet model, we sample it according to the observing strategy to obtain a synthetic data set. We then employ Markov chain Monte Carlo (MCMC) to fit the planetary parameters to the synthetic data, using the No U-Turn Sampler (NUTS) for 1000 warm-up steps, 8000 samples, and 2 chains, as well as a target acceptance probability of 0.9.

\begin{figure}
\includegraphics[width=.9\textwidth]{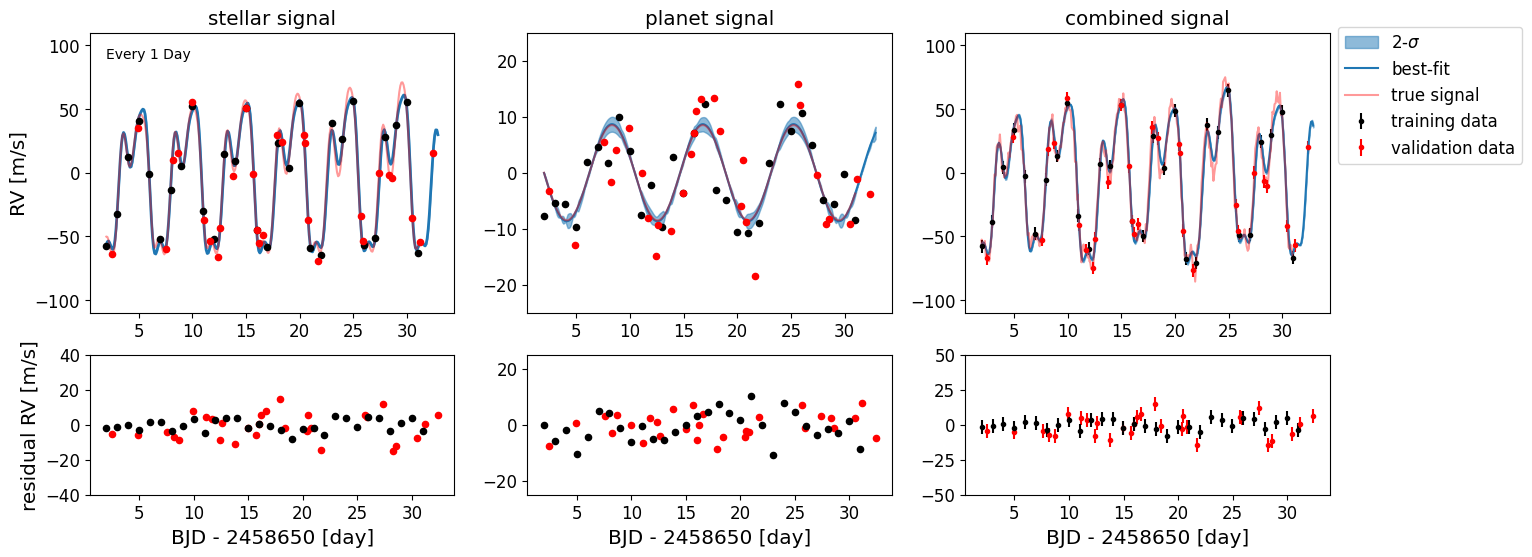}
\includegraphics[width=.9\textwidth]{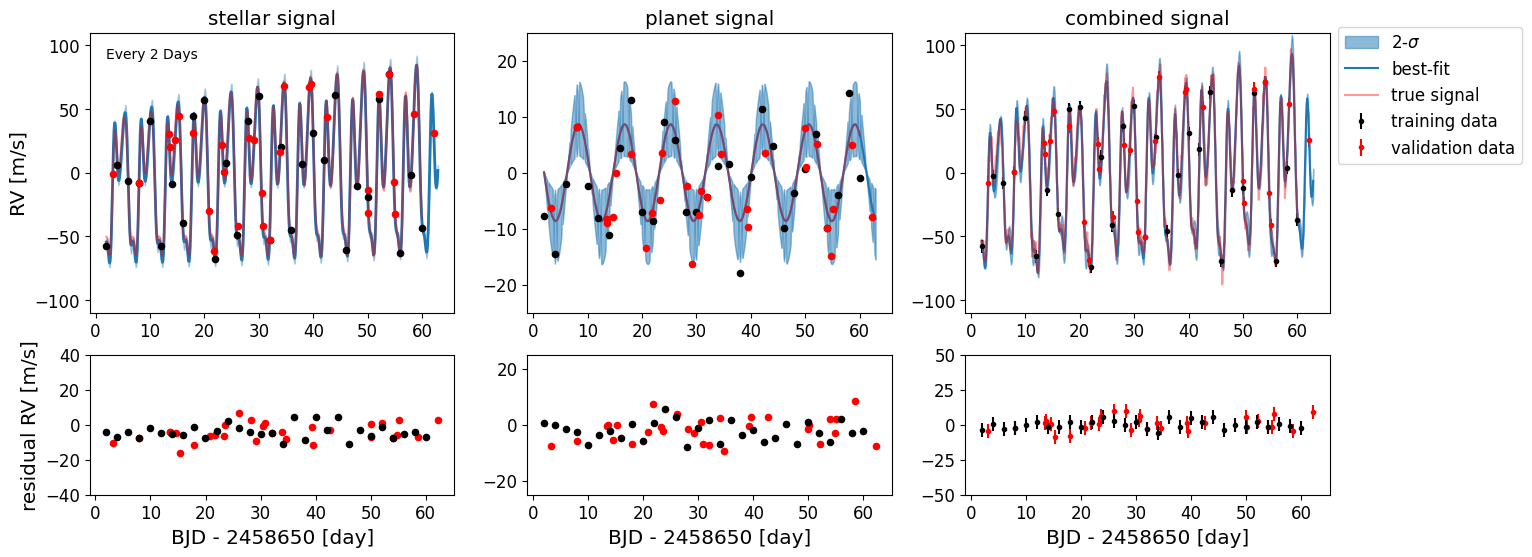}
\includegraphics[width=.9\textwidth]{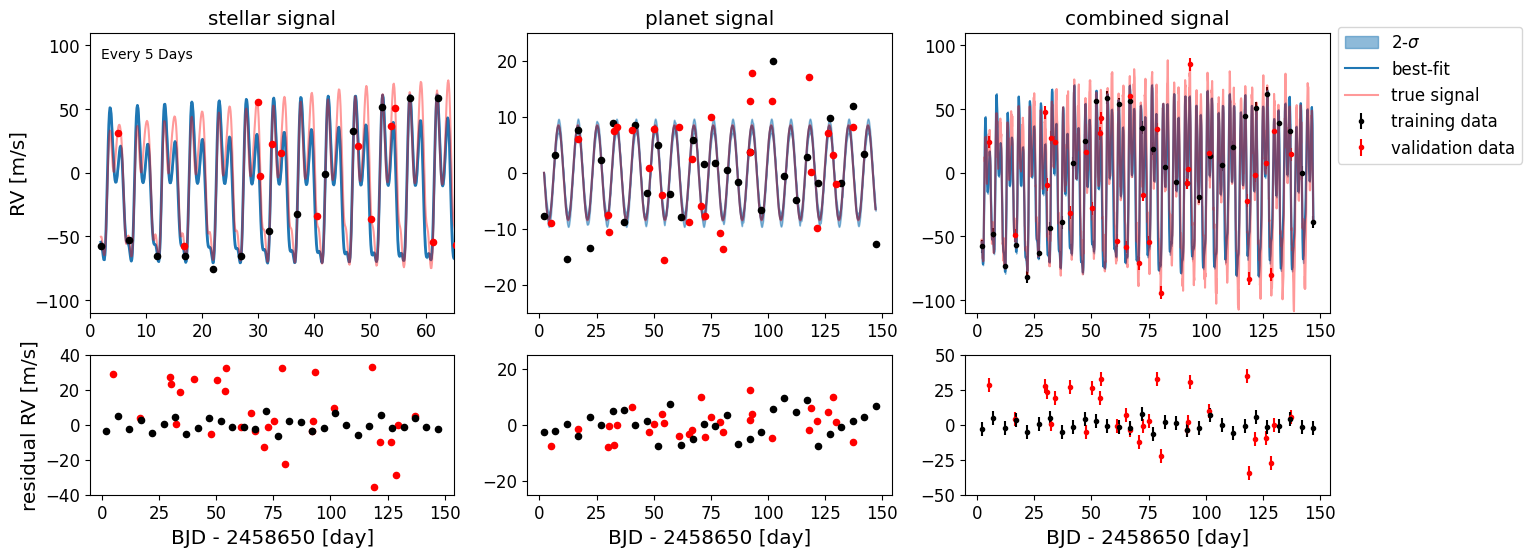}
\caption{MCMC fits for the stellar, planet, and combined signals for three different observing strategies. Blue curves depict the best-fit inferred signals; light red curves illustrate the ``ground truth" signals from which the training (black points) and validation (red points) sets are drawn. The MCMC model is fit only on the training data, which follows the observing strategy. The shorter subplots show the corresponding residuals for each strategy. \textbf{The first set of two rows} correspond to a strategy of observing once per night every night (first row in Table \ref{tab3}). \textbf{The second two rows} correspond to a strategy of observing once per night every other night (second row in Table \ref{tab3}). \textbf{The last two rows} correspond to a strategy of observing once per night every five nights (fourth row in Table \ref{tab3}).}
\label{fig:mcmc-strat}
\end{figure}

In Figure \ref{fig:mcmc-strat}, we show the results of three representative injection-and-recovery tests, corresponding to three observing strategies. We overplot the ground truth model as well as the best-fit inferences from the injection-and-recovery test. For each observing strategy, we also construct a ``control" sample by ``observing" at 30 random times that do not follow the observing schedule, in order to validate whether the model is accurately capturing the overall signal. For each observing strategy, we then consider two sets of residuals. We subtract the ``best-fit" model from both synthetic radial velocity data sets: the one with the prescribed observation times, and the control sampled at random times. We report the standard deviation of these residuals (``Training residual RMS" and ``Validation residual RMS"), the predicted $\sigma_K$ from \gaspery\ (``Fisher Info expected $\sigma_{K}$"), and the uncertainty on the best-fit \textit{K} (``MCMC retrieved $\sigma_{
K}$") in Table \ref{tab3}. For each strategy in Table \ref{tab3}, we average the column values over 10 random observation start times so that the conclusions we reach about generalized strategies are not dependent on the phase sampling of any given individual strategy. The sampled data for the strategies depicted in Figure \ref{fig:mcmc-strat}, meanwhile, all have an observation start time equal to \textit{$T_0$}.

For the strategy of observing once per night every night for 30 nights (first row of Table \ref{tab3}; first two rows of Figure \ref{fig:mcmc-strat}), \gaspery\ predicts an expected value for $\sigma_K$ of 1.4 m/s. An MCMC fit to the same synthetic RV data yields a $\sigma_K$ of +1.3, -1.4 m/s, close to \gaspery's prediction. This strategy allows the model to make an informed fit of both the planet and the stellar signal, resulting in a precise estimate of \textit{K}. 

For the strategy of observing once per night every \textit{other} night until 30 observations are reached (for a baseline of 59 days), \gaspery\ predicts a much larger $\sigma_K$ of 10.4 m/s (second row of Table \ref{tab3}; second pair of rows in Figure \ref{fig:mcmc-strat}). Meanwhile, a fit to the synthetic data sampled at this interval results in a planet RV semi-amplitude uncertainty of +6.8, -5.1 m/s. We find that increasing the uncertainty of the prior on \textit{K} correspondingly increases the $\sigma_K$ of the MCMC fit, and this $\sigma_K$ is never greater than that predicted by \gaspery. This archetype of high-$\sigma_K$ strategy is expected to perform poorly because the sampling falls on a beat frequency between $P_{orb}$ and an integer ratio of $P_{rot}$. These strategies do not necessarily result in large validation set residual spreads, but aliasing from the stellar variability signal hinders our ability to precisely measure the planet's RV semi-amplitude. 

For the strategy of observing once per night every five nights (for a baseline of 245 days), \gaspery\ predicts a $\sigma_K$ of 1.4 m/s, while the MCMC fit yields a very similar $\sigma_K$ of 1.3 m/s (fourth row of Table \ref{tab3}). While this strategy optimizes for $\sigma_K$, the large spread of its validation set residuals and the stellar signal panel in the last row of Figure \ref{fig:mcmc-strat} indicate that this strategy performs poorly at measuring the signal from the star. When we split the simulated data into its stellar and planetary components, we find that the poor accuracy of the best-fit stellar variability signal accounts for almost all of the validation set residual spread. Observing once every five days represents a strategy archetype that is successful at measuring the planet RV signal but is poor at measuring the stellar variability signal because of strong aliasing between the observing cadence and the stellar rotation period (in the case of this strateg
y, the observing cadence is very close to $P_{rot}$, which is 4.86 days). Evidence for this aliasing effect can be seen in the strongly coherent pattern followed by the black observation strategy points, in the stellar signal panel of the last pair of rows in Figure \ref{fig:mcmc-strat}. Observing every five days is therefore representative of a strategy archetype in which sampling the same face of the star over and over again allows observers to more easily separate the stellar signal from the planet signal at the possible expense of relying on an already well-understood stellar signal (see third subplot of Figure \ref{fig:beat-frequency-scenarios}).

\begin{table}
  \centering
  \caption{Retrieved MCMC results for different observing strategies, averaged over 10 random start times}
    \begin{tabular}{ c|c|c|c|c }
    Observation & Fisher Info & MCMC retrieved & Training [m/s] & Validation [m/s] \\ 
     strategy & expected $\sigma_K$ [m/s] & $\sigma_K$ [m/s] & residual RMS & residual RMS \\
     \hline
     Every 1 day$^{\textrm{\textdagger}}$ & 1.4 & +1.3, -1.4 & 3.3 & 7.1\\ 
     Every 2 days$^{\textrm{\textdagger}}$ & 10.4 & +6.8, -5.1 & 2.9 & 7.3 \\  
     Every 2 days, $\sigma_t$=2 hrs & 6.5 & +5.0, -4.4 & 3.1 & 7.0 \\
     Every 5 days$^{\textrm{\textdagger}}$ & 1.4 & +1.3, -1.3 & 3.7 & 16.7 \\ 
In-quadrature & 2.5 & +2.2, -2.2 & 2.6 & 9.6 \\ 
\end{tabular}  
        \tablecomments{Strategies denoted by \textdagger\ are depicted in Figure \ref{fig:mcmc-strat}}
  \label{tab3}
\end{table}

\subsection{Perturbed strategies}
\label{sec:perturbed-strategies}

Results from \citet{burt_simulating_2018} and \citet{gupta_fishing_2023} found that uniform observing strategies were more favorable for optimizing $\sigma_K$ than other classes of strategies because of the greater phase coverage they offered. We therefore investigated whether perturbing the exact-grid time stamps by a spread of 2 hours, or $\sigma_{t=2 hours}$, would improve $\sigma_K$ by distributing observations more uniformly. We explored perturbations with larger spreads, which continued to marginally drive down $\sigma_K$ for the strategy of observing every other day, but these are less realistic to implement.

For the strategy of observing every other day, we find that adding a perturbation of $\sigma_{t=2 hours}$ reduces $\sigma_K$ in both \gaspery\ and the MCMC fit (see third row of Table \ref{tab3}), even though \gaspery\ continues to overestimate $\sigma_K$ relative to the fit. The RV fit residuals for the perturbed strategy remain relatively unchanged from those of the strategy of observing exactly every other day. In practice, the natural variation in observation times for a target over a long baseline may help guard against over-fitting in general. 

\subsection{Observing in-quadrature}
\label{sec:in-quadrature}

\citet{burt_simulating_2018} suggested observing in quadrature (i.e., observing at or near the troughs and peaks of the expected planet RV signal) as a relatively strong (but still sub-optimal) observing strategy for minimizing $\sigma_K$, since this specifically targets the RV semi-amplitude of the planet. We run the same MCMC experiment on a strategy in which we observe in quadrature, where an observation is made at each trough and peak along the planet's Keplerian signal until 30 observations are made. We draw each observation from a Gaussian centered at each extremum, with a spread of 2 hours. We retrieve a $\sigma_K$ from \gaspery\ of 2.5 m/s and a similar $\sigma_K$ from the MCMC fit of 2.2 m/s. Increasing the spread of these observation clusters tends to decrease $\sigma_K$ for both \gaspery\ and the MCMC fit.

Observing in quadrature represents the best-case scenario of a strategy with poor phase coverage, in which observations are clustered only around the peaks and troughs. Such a strategy effectively samples the two most information-rich times in phase. Yet, \citet{burt_simulating_2018} found that more \textit{uniform} observing strategies generally perform better than observing in quadrature. Indeed, we also find that in employing the ``peaks and troughs" strategy, the $\sigma_K$ is higher than that of strategies that sample the phase more uniformly, such as observing once per day.

\section{Discussion} \label{sec:discussion}

In Section \ref{sec:time_sampling}, we set up a framework for evaluating a wide variety of different observing scenarios, with different combinations of strategies and $P_{orb}$ leading to different Fisher Information-derived $\sigma_K$ (also see Figure \ref{fig:cadence_vs_period}). In Section \ref{sec:comparison}, we tested different observing strategies for a specific test-case system, revealing a widely varying range of success at capturing the planet RV semi-amplitude. After generating synthetic planet and stellar variability signals and adding white noise to simulate observed data, we ran a series of MCMC fits on different arrangements of observing times and compared the resulting $\sigma_K$ to \gaspery's predicted $\sigma_K$. These experiments revealed that optimal selection of the observing strategy must account for the planet orbital and stellar rotation periods. In this Section, we discuss how to generalize our findings (Section \ref{sec:ground-rules}), the limitatio
ns to our analysis (Section \ref{sec:limitations}), and potential other use cases in exoplanets (Section \ref{sec:use-cases}). 

\subsection{Ground rules for observation scheduling}
\label{sec:ground-rules}

Examining our findings from the experimental application of \gaspery\ described above, we identify that strategies fall generally into three archetypes. 
\begin{enumerate}
    \item \textbf{Good for planet and star}: strategies for which \gaspery\ predicts a low $\sigma_K$, and an MCMC fit on synthetic data returns low $\sigma_K$ and small residuals on the stellar signal RV fit, e.g., the ``every 1 day'' strategy in Table \ref{tab3}. This strategy provides good sampling of the planet and stellar signals, allowing the data to constrain both simultaneously.
    \item \textbf{Good for planet; bad for star}: strategies for which both \gaspery\ and the MCMC fit on synthetic data give a low $\sigma_K$, but the MCMC fit gives high validation set residuals reflecting poor accuracy on the stellar signal fit. An example is the ``every 5 days'' strategy in Table \ref{tab3}, whose sampling results in only one snapshot of the star's face. As shown in the third subplot of Figure \ref{fig:beat-frequency-scenarios} (corresponding to (5) in Figure \ref{fig:cadence_vs_period}), sampling the same phase of the stellar variability allows us to effectively treat it as just an offset to the planet RV signal, as long as there is no beat frequency between $P_{orb}$ and $P_{rot}$, and the stellar signal is stable and well-understood. \item \textbf{Bad for planet}: strategies for which both \gaspery\ and an MCMC fit on synthetic data yield high $\sigma_K$. These strategies may lie close to a beat frequency of the planet orbital and stellar rotation peri
ods (e.g., the ``every 2 days'' strategy in Table \ref{tab3}), or they may provide poor sampling of the planet RV phase (e.g., the pathological case of observing once per $P_{orb}$, corresponding to (6) in Figure \ref{fig:cadence_vs_period}).  
\end{enumerate}

The observing strategy archetypes listed above assume perfectly spaced time series, but as we show in Section \ref{sec:perturbed-strategies}, the poor performance of some strategies can be mitigated by introducing a small perturbation to the observation times. In general, perturbing information-poor strategies can introduce enough variation to somewhat counter the aliasing effects of beat frequencies, with larger perturbations resulting in smaller $\sigma_K$. We find that increasing the size of perturbations to strategies subject to aliasing by stellar variability in particular tends to decrease $\sigma_K$.

The difference in $\sigma_K$ between observing strategy archetypes is not equally large among all systems. We showed in Section \ref{sec:stellar-parameters} that the variability timescales of a star can affect not only which strategies are optimal, but also whether the available space of optimal strategies is large or small. Stars with spot lifetimes shorter than the rotation period (top right panel of Figure \ref{fig:stellar-parameters-4}) effectively produce a different picture of the spot coverage with each rotation, making it difficult to model the stellar variability and resulting in a higher uncertainty on $K$, regardless of planet orbital period and observing cadence. Differences in $P_{rot}$ affect the sampling needed to achieve good phase coverage of the planet and mitigate the beat frequencies of concern relative to the stellar signal. We therefore emphasize: rotational period is the foremost important timescale to constrain in order to plan an RV follow-up strategy
, followed by the decay timescale for spots ($\tau$). On the other hand, changing correlated noise amplitude ($\sigma_{corr}$) and harmonic complexity ($\eta$) does not affect the optimal choice of strategy itself but rather the coherence of the beat frequencies and therefore the level of care one must place in choosing an optimal strategy. In general, the exact choice of observing strategy and how closely it must be followed matters more for some stars than it does for others.

\subsection{Limitations} \label{sec:limitations}
In this paper we have focused only on single-planet systems. Including additional planet companions in principle simply requires adding three more dimensions ($K$, $T_0$, and $P_{orb}$) to the Fisher Information matrix for each planet. Exploring the larger parameter space introduced by a multi-planet system is beyond the scope of this paper; we direct interested readers to \cite{he_friends_2021} for a discussion of how additional undetected planets typically affect RV campaigns for detected transiting planets. We also chose to limit our exploration of parameter space by considering only zero eccentricity planets, although non-zero eccentricity is possible within the Fisher information framework (see the Appendix in \citet{espinoza-retamal_prospects_2023}). 

One significant limitation of \gaspery\ is that it is reliant on knowing the stellar noise parameters very well. If the wrong stellar noise model is used, \gaspery\ may provide a prediction of $\sigma_K$ that is very different from the retrieved $\sigma_K$ once data is taken. One mitigating factor in the context of \textit{TESS} follow-up is that the sample of host stars with measured rotation periods is predominantly short-period, which have noise timescales that are easier to measure using \textit{TESS} photometry \citep{ricker_transiting_2015}. 

Finally, we make several key assumptions about the noise budget when modeling observations. First, we have assumed that the white noise budget, $\sigma_{WN}$, is comprised of the photon noise and uncorrelated instrument noise. We do not include possible instrument systematics in our correlated noise model. Similarly, any stellar activity component not accounted for in our quasi-periodic Gaussian Process model could likewise affect \gaspery's predicted $\sigma_K$. \Gaspery\ offers users the flexibility to incorporate alternate or additional correlated noise models, using \texttt{tinygp}'s custom kernels.\footnote{A tutorial on folding custom kernels into the \gaspery\ framework can be found at \url{https://github.com/exoclam/gaspery/blob/main/tutorials/custom_kernels.ipynb}.} This flexible framework can also be used to add complexity to the modeled stellar signal itself, such as modifying the quasi-periodic GP kernel to describe multiple spot groups and stellar differential rotation. 

\subsection{Further use cases} \label{sec:use-cases}
In the broader context of scheduling RV follow-up of multiple targets for many different observing programs, an optimization scheme that accounts for the entire instrument queue would not only save human hours in painstakingly scheduling observations, but also enable more efficient resource allocation. \citet{handley_solving_2023} showed that mixed-integer linear programming can generate optimal queue observing schedules for RV follow-up, although they did not model the effect of stellar activity from target hosts on planet RV measurements. One significant challenge to incorporating stellar noise models in such a framework is that not all target hosts' correlated noise are sufficiently -- or even homogeneously -- characterized. 

A Fisher Information approach with GP stellar variability modeling can be generally useful for observations beyond RV measurements to recover exoplanet masses. For example, time-variable stellar heterogeneities pose a challenge to interpreting whether features in transmission spectra come from the planet or host star, which may affect the results of stellar and planetary retrievals \citep{rackham_effect_2023}. In the era of the {\it James Webb Space Telescope} (JWST; \cite{gardner_james_2006}), the problem of scheduling observations around stellar variability may therefore extend beyond RV measurements to include transmission spectroscopy. JWST observations may not be able to capture every Hot Jupiter transit; for particularly short-period planets, then, it may be beneficial to use a \gaspery-like framework to pre-select which transits to observe. Just as we calculated the minimum number of observations required to achieve an uncertainty tolerance in Section \ref{sec:au-mic}, such a framework may also be helpful for calculating the minimum number of observations needed to detect molecules of interest. We leave further investigation of the feasibility of a Fisher Information approach to this problem as a study beyond the scope of this work.

\section{Conclusion} \label{sec:conclusion}
We have presented a generalizable framework, implemented in a pip-installable package called \gaspery, which uses the Fisher Information to design optimal observation strategies for RV follow-up of exoplanets discovered by transit missions like \textit{TESS}. Our implementation of the Fisher Information allows us to calculate the expected RV semi-amplitude uncertainty $\sigma_K$ without simulating the actual data, enabling users to quickly assess the performance of many different strategies at low computational cost. The inclusion of a stellar variability model (and, more generally, our ability to include any noise model characterizable by a Gaussian Process) makes \gaspery\ a flexible tool for planning around correlated noise.

The capabilities of \gaspery\ include:
\begin{itemize}
  \item \Gaspery\ can be used to determine the minimum number of observations required to reach an uncertainty tolerance on the planet RV semi-amplitude in the presence of stellar correlated noise. \Gaspery\ models this correlated noise using a quasi-periodic Gaussian Process kernel and shows that correlated noise causes the relation between $\sigma_K$ and the number of observations, N, to converge to $\sqrt{2/N}$ later than in a case with only white noise.
  \item \Gaspery\ is sensitive to the beat frequencies between the planet orbital period, stellar rotation period, and observation cadence, which can interact to alias the planet RV signal. This results in a large measurement uncertainty on \textit{K}. Whether certain strategies perform well or poorly at constraining \textit{K} depends strongly on the stellar parameters. For example, when the stellar rotation period is less than the spot lifetime of the star, it is easier to capture a stable picture of the stellar signal across phases, making beat frequencies more coherent -- and the choice of observing strategy especially important. 
  \item \Gaspery\ can optimize the information content of an observing strategy given a fixed observing budget and more flexible criteria on how to allocate observations. In the case of an AU Mic-like system with a single planet on a circular orbit, \gaspery\ can be used to produce a diagnostic plot such as Figure \ref{fig:fixed-budget}, showing the expected $\sigma_K$ of a large range of possible observing strategies.
  \item \Gaspery\ is pip-installable, and there are \texttt{Jupyter} notebook tutorials on how to use it to perform the analyses described in this work.\footnote{\url{https://github.com/exoclam/gaspery/tree/main/tutorials}} 
\end{itemize}

In Section \ref{sec:comparison}, we compared \gaspery\ to an MCMC retrieval of $\sigma_K$, finding the two methods in general agreement. Based on these results and the scenarios we illustrated in Figures \ref{fig:cadence_vs_period} and \ref{fig:beat-frequency-scenarios}, we identified the following general principles for the design of RV follow-up observation strategies: 
\begin{itemize}
  \item Strategies with an observing cadence that is either (1) equal or close to the stellar rotation period or (2) a small integer ratio or multiple of $P_{rot}$ will poorly resolve the stellar signal. However, the strategy's ability to measure \textit{K} is relatively independent of this, as long as there is no beat frequency between $P_{orb}$ and $P_{rot}$ and the stellar signal is stable and well-understood.
  \item Strategies that lie on a beat frequency between the stellar rotation period, planet orbital period, and observing cadence will poorly constrain \textit{K}.
  \item Strategies with cadences equal or close to the planet orbital period (or a small integer ratio of $P_{orb}$) will also perform poorly at constraining \textit{K} because they under-sample the planet phase. 
\end{itemize}

In the era of \textit{TESS} and JWST, efficient RV follow-up is more important than ever, and efficient RV scheduling requires consideration of correlated stellar noise. \Gaspery\ addresses this challenge by leveraging our understanding of exoplanet host stars to aid in the planning of RV follow-up observations of their planets. Assuming that target stellar variability timescales are well-known, such a Fisher Information approach to calculating $\sigma_K$ can serve as not only a cost-effective way to compare observing strategies, but also a principled way to methodically identify the best strategy among many.

\section*{Acknowledgments}
We thank Dan Foreman-Mackey, Jiayin Dong, Rae Holcomb, Arjun Savel, Jamie Tayar, Jason Dittmann, Quadry Chance, Sheila Sagear, and Natalia Guerrero for useful discussions that have helped inform and enrich this work. We also thank the anonymous referee for their support in the revision and publication process. This material is based upon work supported in part by the Center for Computational Astrophysics Pre-Doctoral Fellowship program at the Flatiron Institute, as well as the National Science Foundation Graduate Research Fellowship Program under Grant No. 1842473. The Flatiron Institute is a division of Simons Foundation.

We acknowledge that for thousands of years the area now comprising the state of Florida has been, and continues to be, home to many Native Nations. We further recognize that the main campus of the University of Florida is located on the ancestral territory of the Potano and of the Seminole peoples. The Potano, of Timucua affiliation, lived here in the Alachua region from before European arrival until the destruction of their towns in the early 1700s. The Seminole, also known as the Alachua Seminole, established towns here shortly after but were forced from the land as a result of a series of wars with the United States known as the Seminole Wars. We, the authors, acknowledge our obligation to honor the past, present, and future Native residents and cultures of Florida.

\bibliography{final}{}
\bibliographystyle{aasjournal}

\end{document}